\documentclass{wqcd03}                 

\usepackage{mathptmx}                 
\usepackage{epsfig}
\usepackage{xspace}
\usepackage{cite}

\newcommand{\GeV}{\,\,\mathrm{GeV}}
\newcommand{\kbar}{{\bar k}}
\newcommand{\CG}{{\cal G}}
\newcommand{\CGO}{{\cal G}_{\omega}}
\newcommand{\CKO}{{\cal K}_{\omega}}
\newcommand{\as}{\alpha_s}              
\newcommand{\asb}{\bar{\alpha}_s}       
\newcommand{\nf}{n_f}
\newcommand{\eff}{{\mathrm{eff}}}

\newcommand{\gs}{\gamma^{*}}            
\newcommand{\half}{{\textstyle \frac12}}
\newcommand{\kt}{{\bf k}}       
\newcommand{\om}{\omega}
\newcommand{\omhalf}{{\textstyle\frac{\omega}{2}}}
\newcommand{\omc}{\omega_c}
\newcommand{\omp}{\omega_P}
\newcommand{\oms}{\omega_s}
\newcommand{\ord}[1]{\mathcal{O}\left(#1\right)} 

\newcommand{\NLLB}{$\mathrm{NLL_B}$\xspace}

\newcommand{\epjc}[3]{{\it Eur.~Phys.~J.~}{\bf C #1} (#2) #3}
\newcommand{\hep}[1]{{\tt hep-ph/#1}}
\newcommand{\jhep}[3]{{\it JHEP }{\bf #1} (#2) #3}
\newcommand{\jetp}[3]{{\it Sov.~Phys.~JETP }{\bf #1} (#2) #3}
\newcommand{\jetpl}[3]{{\it JETP Lett.~}{\bf #1} (#2) #3}

\newcommand{\npb}[3]{{\it Nucl.~Phys.~}{\bf B #1} (#2) #3}

\newcommand{\plb}[3]{{\it Phys.~Lett.~}{\bf B #1} (#2) #3}
\newcommand{\prd}[3]{{\it Phys.~Rev.~}{\bf D #1} (#2) #3}

\newcommand{\sjnp}[3]{{\it Sov.~J.~Nucl.~Phys.~}{\bf #1} (#2) #3}
\newcommand{\zpc}[3]{{\it Z.~Phys.~}{\bf C #1} (#2) #3}

\confname{QCD@Work 2003 - International Workshop on QCD, Conversano, Italy, 
14--18 June 2003}

\title
{Extending the QCD Perturbative Domain to Higher Energies}

\author{Marcello Ciafaloni\addressmark{a}}

\address[a]{Department of Physics and INFN, Firenze,Italy}

\begin{document}

\begin{abstract}
After a brief introduction to low-$x$ QCD and to resummation approaches, I 
illustrate the predictions for the gluon Green function and splitting function
of a recent renormalization--group improved small-$x$ resummation scheme. I 
argue, on this basis, that the range of validity of perturbative calculations 
is considerably extended in rapidity with respect to leading log expectations.
The perturbative high-energy exponents are predicted in a phenomenologically 
interesting range, and significant preasymptotic effects are found. In 
particular, the splitting function shows a shallow dip in the moderate-$x$ 
region, followed by the expected small-$x$ power increase.
\end{abstract}

\maketitle

\section{Low-$x$ Physics in QCD}

The description of high-energy hard cross--sections in QCD perturbation theory
has a long history. Starting from the early BFKL prediction of rising cross 
sections~\cite{BFKL} at leading log$s$ level (LL) and from the 
$k$-factorization framework~\cite{LOImpact}, up to the difficult calculation 
of the
next--to--leading (NLL) kernel~\cite{NLLFL,NLLCC} and to its improvements
~\cite{Salam1998,CC,CCS1,ABF2000,ABF2001,THORNE,SCHMIDT,BFKLP}, a lot of work 
has been devoted to fill the gap between theory and small-$x$ phenomenology.

On the experimental side, the DIS structure functions at HERA~\cite{HERA} show
a rise which is well described by large scaling violations arising in low order
 DGLAP~\cite{DGLAP} evolution. On the other hand, data for two--scale processes
like $\gs(Q) - \gs(Q_0)$ for which the DGLAP picture is inappropriate, seem 
incompatible with the strong rise predicted by LL BFKL evolution. There is 
thus a need to reconcile the two theoretical pictures, which actually differ 
by the order of resummation of large logarithms (log~$Q^2$ vs. log~$s$) in 
the perturbative series.

On the theoretical side, one expects QCD perturbation theory at low $x$ to be 
governed by some effective coupling $\as(\kt^2)\log 1/x$ which risks to be 
strong for two independent reasons. Firstly, because of the logarithmic 
enhancements due to the opening of the small$x$ gluon phase space. This 
motivates the calculation of next-to-leading corrections to the hard-Pomeron 
intercept, which however turn out to be quite large and negative~\cite
{NLLFL,NLLCC}, pointing to some instability of the log~$s$ hierarchy. 
Secondly, because  energy evolution -- by diffusion or tunneling~\cite{CCSS2}
-- may give rise to low values of $\kt$, the gluon momentum transfer, thus 
implying a drift towards a really non-perturbative asymptotic  regime~
\cite{JKCOL,Lipatov86, CC97, CC1}, characterized by soft physics.
We must control, therefore, higher order subleading terms, and analyze the 
perturbatively accessible rapidity range.
 
A tentative answer to the above question marks about the perturbative 
behaviour comes from the (double) resummation approaches. The idea is to take 
into account higher order subleading terms in both variables (log~$s$ and 
log~$Q^2$), based on some distinctive features of such terms. For instance, 
the ``duality'' approach~\cite{ABF2000, ABF2001} starts from the approximate 
symmetry between scaling violations and energy dependence in order to hint 
at a resummed form of the gluon splitting function. The ``renormalization-group
improved'' (RGI) approach~\cite{Salam1998,CC,CCS1} starts from 
BFKL evolution at NLL level, and resums those higher order subleading terms 
which are required by the RG behaviour for both $Q\gg Q_0$ and $Q_0\gg Q$. 
Other 
resummation models are available also~\cite{THORNE,SCHMIDT,BFKLP}.
The RGI approach is able to provide predictions~\cite{CCSS3} for both the 
gluon splitting function and the gluon density itself, and I will concentrate 
on that approach in the following.  

 The BFKL equation is, to start with, an evolution equation in the rapidity 
$Y=\log(s/QQ_0)$, induced by a kernel $K(\kt, \kt')$ in $\kt$ --space, which 
can be expanded in $\as$ and has been calculated up to second order. The LL 
level is dominated by (effective) ladder diagrams with gluon exchange and the 
corresponding kernel is scale--invariant and given by ($\asb=N_c\as/\pi$).
\begin{equation}\label{LLke}
\as K_0(\kt, \kt') = \frac {\asb}{(\kt-\kt')^2}~+~virtual~ terms.
\end{equation}
The highest eigenvalue of $K_0$ yields the exponent for the energy dependence
and is $\oms = 4\log2\asb$, a value which is too large compared to HERA data 
(e. g., 0.55 for $\as$=0.2, compared to an observed value of about 0.2).
Similarly, the gluon anomalous dimension, given by $\asb/\om$ in $\om =N-1$
moment space at one--loop level, acquires higher order singularities with 
positive residues. Roughly speaking, the LL approximation grossly overestimates
both two--scale cross--sections' rise and scaling violations.

The NLL kernel, extracted in~\cite{NLLFL,NLLCC} after many years of work of 
various groups, yields some qualitatively new features and question marks.
Firstly, the corrections to the high--energy exponent are negative (which is 
good), but with a quite large coefficient, as follows
\begin{equation}\label{LLexp}
\oms=\oms^0~(1 - 6.47~\asb~+~...).
\end{equation}
This points towards an instability of the leading--log~$s$ hierarchy, which is
confirmed by the risk of oscillatory behaviour of the gluon density~\cite
{ROSS98} for sizeably different external scales. Secondly, running coupling
effects are taken into account, and the scale of the coupling suggested by
the kernel is $(\kt-\kt')^2~=~q^2$. Therefore, in order to avoid divergent 
integrations, we need to regularize the Landau pole, by introducing some cutoff
(or freezing parameter) $\bar{k}$. However, when the energy increases, there 
is a favoured evolution towards smaller values of $\kt~=~\ord{\bar{k}}$, by
diffusion or tunneling, and this implies that the asymptotic high--energy
exponent becomes $\bar{k}$-dependent, i.e., non--perturbative (asymptotic
Pomeron). Therefore, NLL calculations raise more questions than they are able
to answer, and call for the inclusion of higher order terms.

\section{The Renormalization Group Improved Approach}

Higher order subleading contributions are not known in detail. We can argue, 
however, that a class of them are large, and required in order to build 
single--logarithmic scaling violations when either $Q$ or $Q_0$ is the leading
scale~\cite{Salam1998}, and $Q^2/s$ or $Q_0^2/s$ is the corresponding Bjorken
variable. Taking into account such terms is equivalent to resumming subleading
terms in log~$s$ which are known because they are leading in log~$Q^2$ and are 
provided therefore by the renormalization group. This idea was used in order 
to build the RGI approach of~\cite{CCS1}, where the solution of the homogeneous
small-$x$ equation was studied in detail by the use of the $\om$--expansion
method~\cite{CC} so as to calculate stable high--energy exponents and gluon 
anomalous dimension. In recent papers~\cite{CCSS3} the approach has 
been extended to the full inhomogeneous equation in a slightly different
resummation scheme, which is more suitable for numerical evaluation. I am thus
able to describe results for both the two--scale gluon density and for the
splitting function in the collinear limit. Results for the splitting function 
have been recently obtained in the duality approach also~\cite{ABF2003}.

The basic problem considered in~\cite{CCSS3} is the calculation of the
(azimuthally averaged) gluon Green function $G(Y;k,k_0)$ as a function
of the magnitudes of the external gluon transverse momenta $k \equiv
|\kt|,\;k_0 \equiv |\kt_0| $ and of the rapidity $Y\equiv \log
\frac{s}{k k_0}$. This is not yet a hard cross section, because we
need to incorporate the impact factors of the
probes~\cite{LOImpact,BaCoGiKy,Bartels02}.
Nevertheless, the Green function exhibits most of the physical
features of the hard
process, if we think of $k^2,\;k_0^2$ as external (hard) scales. The
limits $k^2\gg k_0^2$ ($k_0^2 \gg k^2$) correspond conventionally to
the ordered (anti-ordered) collinear limit. By definition, in the
$\om$-space conjugate to $Y$ (so that $\hat{\om} = \partial_Y$) we
set
\begin{equation}
 \label{defGGF}
 \CGO(\kt,\kt_0) ~\equiv [\om - \CKO]^{-1} (\kt,\kt_0) \,, 
\end{equation}
and
 \begin{equation}\label{eqGGF}
 \om \CGO(\kt,\kt_0) ~= \delta^2(\kt-\kt_0) +
 \int d^2 \kt' \; \CKO(\kt,\kt') \CGO(\kt',\kt_0) \;,
\end{equation}
where $\CKO(\kt,\kt')$ is a kernel to be defined, whose $\om = 0$
limit is related to the BFKL $Y$-evolution kernel discussed before.

The precise form of the kernel $\CKO$ is given in Ref.~\cite{CCSS3}
However, the basic features of the RGI approach is illustrated by
a simple observation: in BFKL
iteration, all possible orderings of transverse momenta are to be
included, the ordered (anti-ordered) sequence
$k \gg k_1 \cdots \gg k_n \cdots \gg k_0$
($k \ll k_1 \cdots \ll k_n \cdots \ll k_0$)
showing scaling violations with Bjorken variable $k^2/s$ ($k_0^2/s$).
Therefore, if only leading $\log k^2$ contributions were to be
considered, the kernel $\CKO$ acting on
$\frac{1}{k^2}\left(\frac{k^2}{k_0^2}\right)^\gamma$
would be approximately represented by the
following eigenvalue function (in the frozen coupling limit)
\begin{eqnarray}\label{CKOform}
 \frac{1}{\om} \CKO \rightarrow \asb
 \left( \frac1{\gamma+\omhalf} + \frac1{1+\omhalf-\gamma} \right)
 \left( \frac1{\om} + A_1(\om) \right) + \cdots \nonumber&& \\ 
 \asb \equiv \as \frac{N_c}{\pi} \;, \hspace{60mm}&&
\end{eqnarray}
where $\gamma^{(1)}_{gg} = \asb \left( \frac1{\om} + A_1(\om) \right)$
is the one-loop gluon-gluon anomalous dimension and we have introduced
the variable $\gamma$ conjugate to $\log k^2$. Note, in fact, that
Eq.~(\ref{CKOform}) reduces to the normal DGLAP evolution~\cite{DGLAP}
in $\log k^2$ ($\log k_0^2$) in the two orderings mentioned before,
because $\gamma+\omhalf$ ($1+\omhalf-\gamma$) is represented by
$\partial_{\log k^2}$ ($\partial_{\log k_0^2}$) at fixed values of $x
= k^2 /s$ ($x_0 = k_0^2 /s$) in the ordered (anti-ordered) momentum
region. Note also the $\om$-dependent shift~\cite{Salam1998,CC,CCS1}
of the $\gamma$-singularities occurring in Eq.~(\ref{CKOform}), which
is required by the change of scale ($k_0^2$ versus $k^2$) needed to
interchange the orderings, i.e., $x_0$ versus $x$.

We thus understand that the $\om$-dependence of $\CKO$ is essential
for the resummation of the collinear terms and can be used to
incorporate the exact LL collinear behaviour, while on the other hand,
the $\om \to 0$ behaviour of $\CKO$ is fixed by the BFKL limit up to
$\ord{\om}$ terms, so as to incorporate exact LL and NLL kernels. Such
requirements fix the kernel up to contributions that are NNL in $\ln
x$ and NL in $\ln Q^2$. The resulting
integral equation to be solved by the definition~(\ref{eqGGF}) is
thus a running coupling equation with non linear dependence on $\as$
at appropriate scales, and it has a somewhat involved $\om$-dependence
in the improved kernels $\CKO$. Its solution has been found
in~\cite{CCSS3} by numerical matrix evolution methods in $k$- and $x$-
space~\cite{BoMaSaSc97}, where the typical $\om$-shifted form in the
example~(\ref{CKOform}) corresponds to the so-called consistency
constraint~\cite{Ciaf88,LDC,KMS1997}. Furthermore, introducing the
integrated gluon density
\begin{eqnarray}\label{gluonDensity}
 x g(x,Q^2) &\equiv& \int^{Q^2} d^{2\!}\kt \; 
 G^{(s_0=k^2)}(\log 1/x; |\kt|, k_0) \;, \\
 \CG_\om^{(s_0=k^2)}& \equiv& \left(\frac{k}{k_0}\right)^\om \CGO \;,
\end{eqnarray}
the resummed splitting function $P_\eff(z,Q^2)$ is defined by the
evolution equation
\begin{equation}\label{splitting}
 \frac{\partial g(x,Q^2)}{\partial \log Q^2} = \int \frac{dz}{z}\;
 P_\eff(z,Q^2)\; g\left( \frac{x}{z}, Q^2 \right)\,,
\end{equation}
and has been extracted~\cite{CCSS3} by a numerical deconvolution
method~\cite{CCS3}. We note immediately that $P_\eff$ turns out to be
independent of $k_0$ for $Q^2 \gg k_0^2$, yielding an important check
of RG factorisation in our approach.
\section{Gluon Green's Function}

\begin{figure}[h!]
  \centering
  \includegraphics[width=0.4825\textwidth]{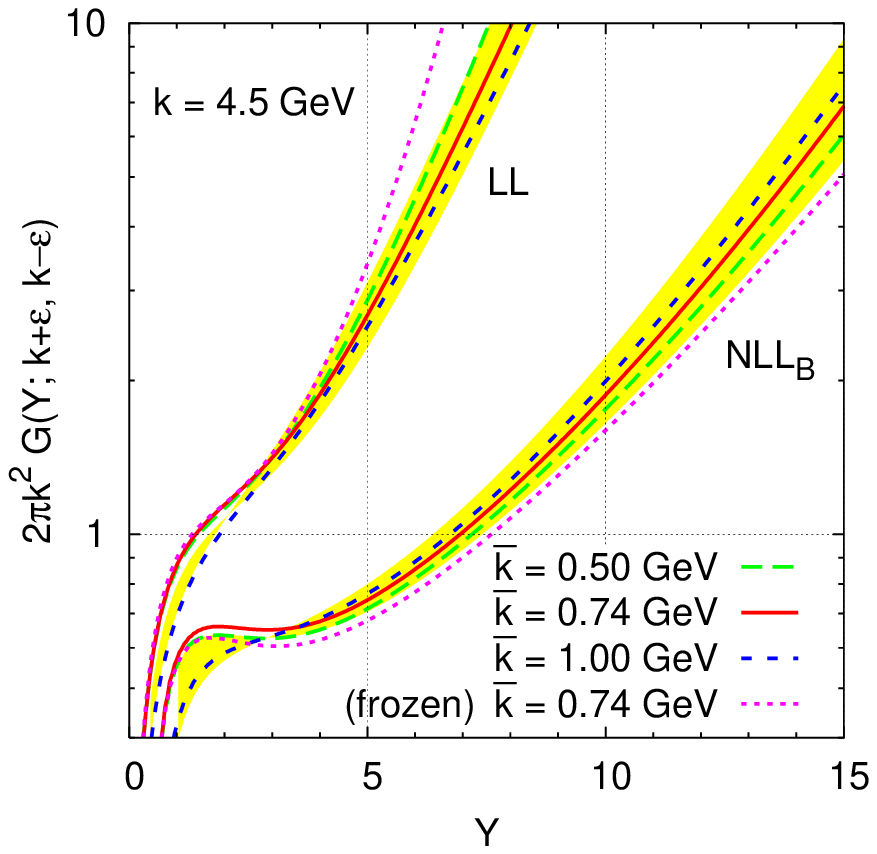}\hfill
  \includegraphics[width=0.50\textwidth]{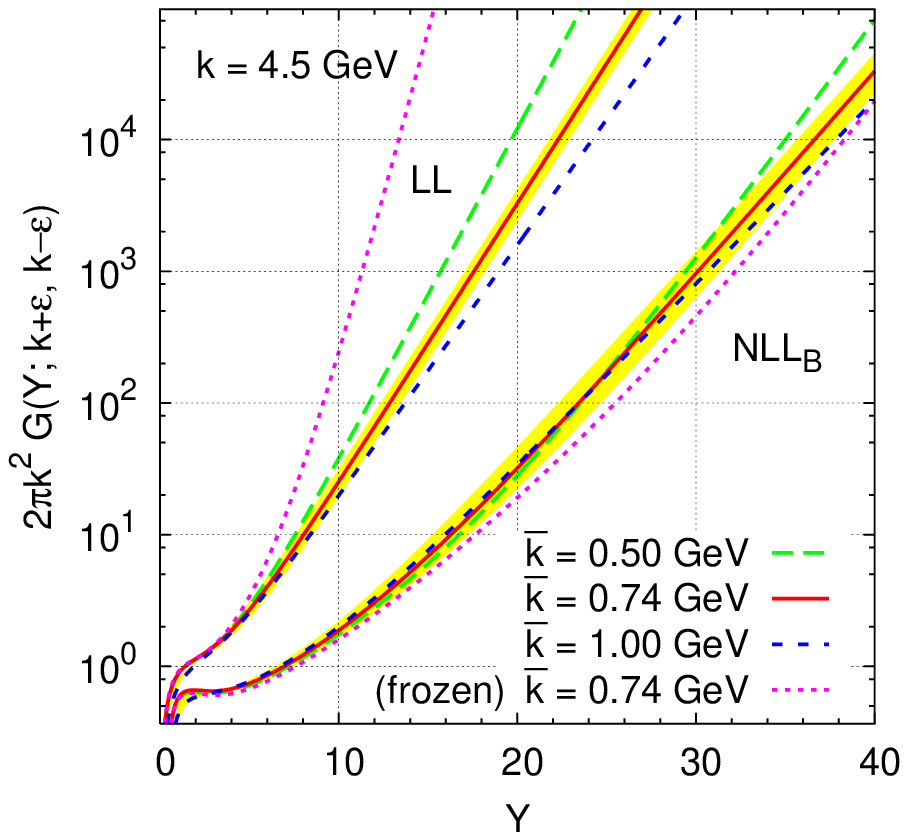}
  \caption{\it Green's function calculated with four different
    infrared regularisations of the coupling, shown for LL and RGI NLL
    (`NLL$_\mathrm{B}\!$') evolution. The bands indicate the
    sensitivity of the $\kbar=0.74\GeV$ results to a variation of
    $x^2_\mu$ in the range $\frac12$ to $2$. The left and right hand
    plots differ only in their scales.
  }
  \label{f:manyReg}
\end{figure}

Results for $G(Y;k,k)$%
\footnote{Actually,slightly different values of the scales in $G$ are taken, 
  namely $G(Y;k+\epsilon,k-\epsilon)$ with $\epsilon
  = 0.1 k$, in order to avoid sensitivity to the discretisation of the
  $\delta$-function initial condition in \ref{eqGGF} (cf.\ 
  Ref.~\cite{CCSS3} for a detailed discussion).}  
are shown in Fig.~\ref{f:manyReg}.
In addition to the solution based on our RGI approach (\NLLB) the
figure also has `reference' results for LL evolution with kernel
$\asb(x_\mu^2 q^2) K^0_0$. The one-loop running coupling (with
$\nf=4$), is regularised either by setting it to zero below a
scale $\kbar$ (`cutoff') or by freezing it below that scale
($\as(q^2<\kbar^2) = \as(\kbar^2)$). The cutoff regularisation is supposed
to be more physical since it prevents diffusion to
arbitrarily small scales and is thus more consistent with confinement --
accordingly three cutoff regularisations 
are shown and only one frozen regularisation. The $\kbar =
0.74\GeV$ cutoff solution is presented together with an uncertainty
band associated with the variation of $x^2_\mu$ between $\frac12$ and
$2$, $x_\mu$ being a renormalization--scale testing parameter, such that
$\asb(q^2)$ is replaced by $\asb(x_\mu^2q^2) + b~\asb^2~log~x_\mu^2$ in the 
part of the kernel linear in $\as$.

Solutions of (\ref{eqGGF}) with an IR-regularised coupling generally
have two domains~\cite{JKCOL,Lipatov86,CC97,CCSS2},
separated by a critical rapidity $Y_c(k^2)$. For the
intermediate high-energy region $1 \ll Y < Y_c(k^2)$, one
expects the perturbative `hard Pomeron' behaviour with exponent $\oms$,
\begin{equation}\label{Gpert}
 k^2 G(Y;k,k) \sim \frac1{\sqrt{Y}}
 \exp \big[ \oms(\as(k^2)) Y + \Delta(\as, Y) \big] \;,
\end{equation}
and diffusion corrections~\cite{KOVMUELLER,ABB,LEVIN,CMT}
parametrised by $\Delta(\as, Y)$. Beyond $Y_c$, a
regularisation-dependent non-perturbative `Pomeron' regime takes over
\begin{equation}\label{Gnonpert}
 k^2 G(Y;k,k) \sim \left(\frac{\kbar^2}{k^2}\right)^{\xi} e^{\omp
   Y}\,,\qquad 
   \begin{array}{rl}
     \mathrm{LL}\;\,\,: & \xi = 1 \\ 
     \mathrm{NLL}_B: & \xi = 1 + \omp 
   \end{array}
\end{equation}
where the non-perturbative exponent $\omp$ satisfies~\cite{CCS1} $\omp\sim
\oms(\as(\kbar^2))$ and hence
is formally larger than $\oms(\as(k^2))$.\footnote{The behaviour
  \ref{Gnonpert} with $\omp > \oms(\as(k^2))$ is a general feature
  of linear evolution equations such as \ref{eqGGF}, but not of actual
  high energy cross sections, which are additionally subject to
  non-linear effects and confinement.} 

The value of $Y_c$ depends strongly on $k$. In the tunnelling
approximation, it can be roughly estimated by equating
eqs.~(\ref{Gpert}) and (\ref{Gnonpert}) to yield~\cite{CCS2,CCSS2},
for any given regularisation procedure,
\begin{equation}\label{Yc}
 Y_c(k^2) \simeq \frac{\xi\log(k^2/\kbar^2)}{\omp - \om_s(\as(k^2))} \;,
\end{equation}
(again with $1+\omp \to 1$ for LL), showing an approximately linear increase
of $Y_c$ with $\log k^2$.

Within this logic, several aspects of Fig.~\ref{f:manyReg} are worth
commenting. The most striking feature of the LL evolution is its
strong dependence on the non-perturbative regularisation, even for
rapidities as low as $5$. The exact value of $Y_c$ depends on the
regularisation being used, ranging between $5$ and $10$.
In contrast, \NLLB evolution remains under perturbative control up to
much larger rapidities and the NP pomeron behaviour takes over only
for $Y > 25$, where the three cutoff solutions start to
diverge\footnote{Renormalisation scale uncertainties of the
resummed results are sizeable -- of the order of several tens of
percent for $Y>4$ -- but seem anyhow quite modest compared to the
order(s) of magnitude difference with LL. The $x_\mu$
  dependence of the LL solution is somewhat smaller than for \NLLB --
  this may seem surprising, but at larger $Y$ the LL solution is in
  the NP domain, where non-linearities (in $\as$) reduce the $x_\mu$
  dependence.}.
Therefore, $Y_c$ is considerably larger for the resummed evolution,
as consequence of the fact that subleading corrections lower both the
PT and -- even more -- NP exponents (see Eq.\ref{Yc}).
 
We note that at
their respective $Y_c$'s the \NLLB Green's function is an order of
magnitude larger than the LL one: the subleading corrections increase
the overall amount of BFKL growth remaining within perturbative
control. However, large densities $G\sim 1/\as$ are reached at cosiderably 
larger values of $Y$, so that saturation effects~\cite{KOVMUELLER} are 
pushed towards higher energies. For the reference values $k=4.5$ GeV,
$\as\sim 0.2$ this translates to Y of order 15, close to the kinematical limit 
of LHC, as a very rough estimate.

\begin{figure}[h!]
  \newcommand{\figfrac}{0.47} \centering
  \includegraphics[height=\figfrac\textwidth]{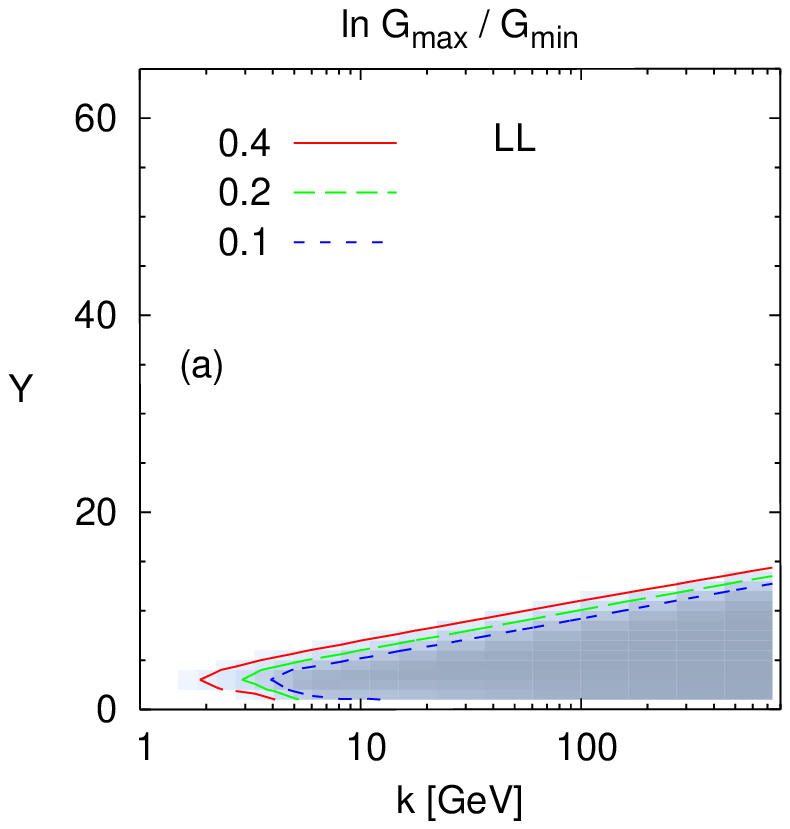} \hfill
  \includegraphics[height=\figfrac\textwidth]{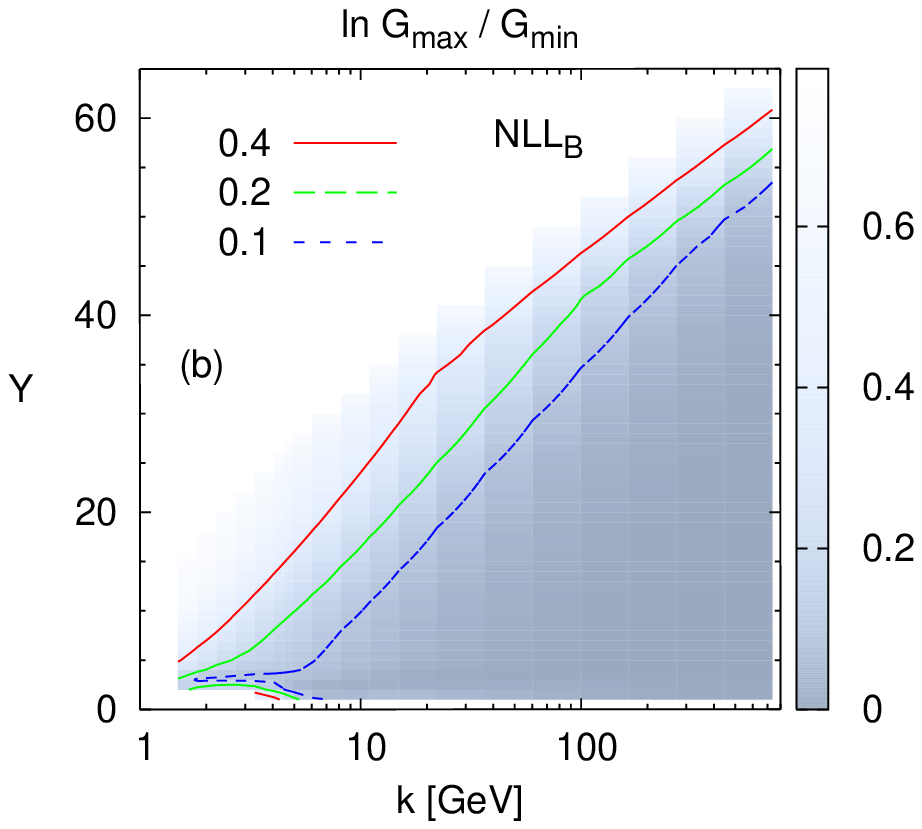}
  \caption{\it Contour plots showing the sensitivity of
    $G(Y,k+\epsilon, k-\epsilon)$ to one's choice of non-perturbative
    regularisation, as obtained by examining the logarithm of the ratio
    of the regularisations giving the largest and smallest result for
    $G$. Darker shades indicate insensitivity to the NP
    regularisation, and contours have been drawn where the logarithm
    of the ratio is equal to $0.1$, $0.2$ and $0.4$. Plot (a) shows
    the result for LL evolution, while (b) shows RGI NLL evolution
    (\NLLB). The regularisations considered are those of
    Fig.~\ref{f:manyReg}.}
 \label{f:valRegions}
\end{figure}

In Fig.~\ref{f:manyReg} only a single value of $k$ is considered.
The question of NP contributions is summarised more generally in
Fig.~\ref{f:valRegions}, which shows contour plots of the logarithmic
spread of the four regularisations. Darker regions are less IR
sensitive, and contours for particular values of the spread have been
added to guide the eye. Here too one clearly sees the much larger
region (including most of the phenomenologically interesting domain)
that is accessible perturbatively after accounting for subleading
corrections.

So let us now therefore return to Fig.~\ref{f:manyReg} and examine the
characteristics of the \NLLB Green's function in the perturbatively
accessible domain, which should be describable by an equation of the
form (\ref{Gpert}). The first feature to note is that the growth
starts only from $Y\sim 4$. This suggests that at today's
collider energies (implying $Y< 6$~\cite{OPAL,L3}), it will at
best be possible to see only the start of any growth. This
preasymptotic feature is partly due to the slow opening of small-$x$
phase space~\cite{NONASYMP} implicit in our $\om$--shifting procedure.

Once the growth sets in,
the issue is to
establish the value of $\oms$ appearing in (\ref{Gpert}).  This is a
conceptually complex question because in contrast to the
fixed-coupling case, $\oms$ no longer corresponds to a Regge
singularity.

There are running-coupling diffusion corrections
$\Delta(\as, Y)$~\cite{KOVMUELLER,ABB,LEVIN,CMT}, whose leading
contribution, $\sim Y^3$,  for this model is~\cite{CCSS3}
\begin{equation}\label{diffCorr}
 \Delta(\as, Y) \simeq  \frac{Y^3}{24}
 \left[ \frac{\partial}{\partial \log k^2}\oms(\as(k^2)) \right]^2
 \chi_\eff ''(\half)\,.
\end{equation}
In addition, $\Delta(\as, Y)$ contains terms with weaker $Y$
dependences, including $Y^2$ and $Y$.  Such terms can be disentangled
by the method of the $b$-expansion~\cite{CCSS1}.\footnote{ In the
  $b\to0$ limit, with $\as(k^2)$ kept almost fixed, the
  non-perturbative Pomeron is exponentially
  suppressed~\cite{CCSS1,ABF2001}, so that the
  $b$-expansion can also be used as a way of defining a purely
  perturbative Green's function without recourse to any
  particular infrared regularisation of the coupling~\cite{CCSS1}.}.
\begin{figure}[h!]
\centering
\includegraphics[height=0.48\textwidth]{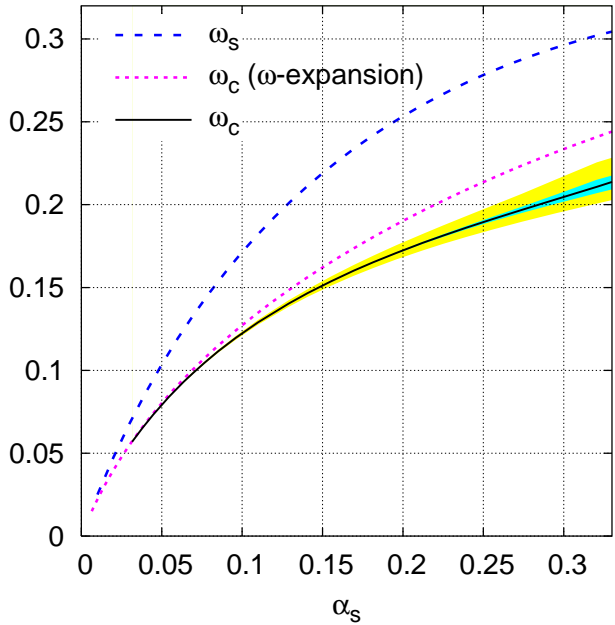}
\caption{\it The small-$x$ exponents: the Green's function effective exponent
  $\oms$ is shown to first order in the $b$-expansion; the splitting
  function exponent $\omc$ is shown together with NP and
  renormalisation scale uncertainty bands, defined in
  figure~\ref{f:Peff}. Also shown, for reference, is the result for
  $\omc$ using the method of~\cite{CCS1}, for $b(\nf = 4)$.}
\label{f:omc}
\end{figure}%
Since running-coupling diffusion corrections start only at order
$b^2$, it is possible to give an unambiguous definition of $\oms$ up
to first order in $b$, while retaining all orders in $\as$ for non
running-coupling effects. The result is shown in Fig.~\ref{f:omc} as a
function of $\as$ and as has been found in previous work~\cite{CCS1},
there is a sizeable decrease with respect to LL
expectations.\footnote{A direct comparison with earlier
  results for $\oms$~\cite{CCS1} is not possible, because they are based on a
  different definition (the saddle-point of an effective
  characteristic function), which is less directly related to the
  Green's function. Nevertheless, the present results are consistent with
  previous ones to within NNLL uncertainties.} %
Furthermore, the leading diffusion corrections in (\ref{diffCorr})
turn out to be numerically small, about an order of magnitude down
with respect to the LL result, due to a sizeable decrease of the
diffusion coefficient $\chi_\eff''$, over and above the decrease
already discussed for $\oms$.

\section{Resummed Splitting Function}
\begin{figure}[ht]
\centering\includegraphics[width=0.50\textwidth]{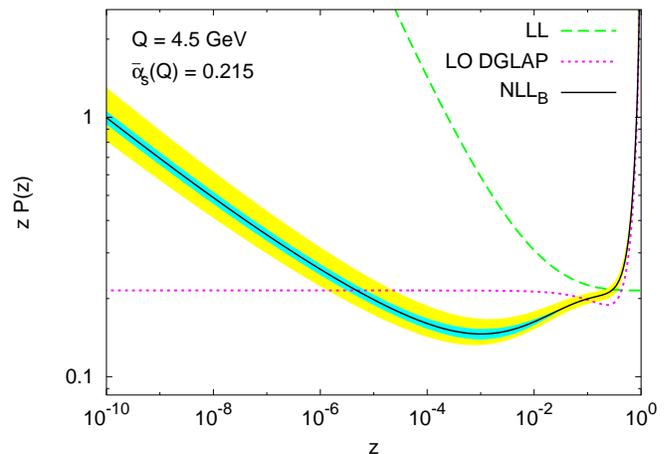}
\caption{\it Small-$x$ $\mathrm{NLL_B}$ resummed splitting function,
  compared to the pure 1-loop DGLAP and the (fixed-coupling) LL BFKL
  splitting functions. The central $\mathrm{NLL_B}$ result corresponds
  to $x_\mu=1$, $\kbar=0.74\GeV$; the inner band is that obtained by
  varying $\kbar$ between $0.5$ and $1.0\GeV$, while the outer band
  corresponds to $\frac12 < x_\mu^2 < 2$.}
\label{f:Peff}
\end{figure}

The Green's function $G(Y;k,k_0)$ has been investigated in the collinear
limit $k\gg k_0 \sim \kbar$ also. In such a case the sensitivity to the
IR regularization is much stronger.
However, many arguments in the BFKL
framework,\cite{Lipatov86,JKCOL,CC1,CCS1,ABF2001,CCS3}, have been
given in favour of {\it factorisation}, Eq.~\ref{splitting}, with
the small-$x$ splitting function $P_{gg}(z,Q^2)$ being independent of
the IR regularisation.  The most dramatic demonstration of
factorisation is perhaps in the fact that a numerical extraction of
the splitting function from the Green's function by deconvolution
gives almost identical splitting functions regardless of the
regularisation. This is illustrated in Fig.~\ref{f:Peff}, 
where the
solid line and its inner band represent the result of the
deconvolution together with the uncertainty resulting from the
differences between the three cutoff regularisations.
The resulting regularisation dependence is pretty small, and at higher
$Q$ it diminishes rapidly as an inverse power of $Q$, as expected from
a higher twist effect.

Several features of the resummed splitting function are worth
commenting and comparing with previous NLL calculations, including
various types of resummations~\cite{CCS1,ABF2001,THORNE}. Firstly, at
very large-$x$ it approaches the normal DGLAP splitting function. The
momentum sum rule is satisfied to within a few parts in $10^4$. At
moderately small-$x$ the splitting function is quite strongly
suppressed with respect to the LL result and shows not a power growth,
but instead a significant dip (of about $30\%$ relative to the LO
DGLAP value, for $\asb = 0.215$). Dips of
various sizes and positions have been observed before
in~\cite{THORNE,ABF2000,CCS3} though this is significantly shallower
than that found in~\cite{THORNE} at NLL order and similar to that
found in the $\om$-expansion~\cite{CCSS3} and in the duality approach
~\cite{ABF2001,ABF2003}.  

At very small-$x$ one finally sees the BFKL growth of the splitting
function. We recall that the branch cut, present for a fixed
coupling, gets broken up into a string of poles, with the rightmost
pole located at $\omc$, to the left of the original branch point
($\oms$), $\oms - \omc \sim b^{2/3} \as^{5/3}$~\cite{CCS1}. The origin
of this correction is similar to that of the $b^{2/3} \as^{5/3}$
contributions to $\omp$ for cutoff regularisations~\cite{HR92}. The
dependence of $\omc$ on $Q$ is shown in Fig.~\ref{f:omc} together with
its scale and IR regularisation dependence. It is slightly lower than
the earlier determination in the
$\om$-expansion~\cite{CCS1}\footnote{when compared with the same
  flavour treatment ---
  the value of $\omc$ in Fig.~6 of~\cite{CCS1} actually refers to $b(\nf = 0)$,
  while that of Fig.~\ref{f:omc} here is for $b(\nf = 4)$.}.  Both
determinations are substantially below $\oms$, as expected.

To sum up, the resummed gluon density at comparable scales $k\sim k_0$ stays
perturbative for a wide rapidity range - including most of the phase space 
available at next generation colliders - due to a suppression of the
non-perturbative Pomeron and of diffusion corrections~(Fig.\ref{f:valRegions}).
The expected increase with energy starts slowly~(Fig.\ref{f:manyReg}), due 
to small-$x$ phase space effects, and is regulated by the exponent 
$\oms$ (Fig.~\ref{f:omc}) which, in the relevant $Q$-range is roughly of size
$\as(Q^2)$. In the collinear region
$k\gg k_0$ the gluon splitting function can be factored out, shows a shallow 
dip in the moderate small-$x$ range~(Fig.\ref{f:Peff}), and its increase is 
regulated by the exponent $\omc < \oms$, also shown in Fig.~\ref{f:omc}.

A realistic prediction of cross-sections requires the inclusion of impact
factors along the lines of~\cite{CCSS3}, and of the quark sector along the 
lines of~\cite{CC1,CCS1}.
 But we can already say that resummed results show 
interesting preasymptotic effects and are, very roughly, closer to low 
order predictions than expected. This in turn may provide a preliminary 
explanation of the apparent smoothness of small-$x$ cross-sections despite
the occurrence, in their description, of large perturbative coefficients
and of various strong-coupling phenomena.

\section*{Acknowledgements}
 I wish to thank Dimitri Colferai, Gavin Salam and Anna Stasto for a number of 
discussions and suggestions on the topics discussed here. I also thank 
Guido Altarelli 
and Stefano Forte for various conversations on similarities and differences of
the duality approach. I am grateful to the workshop organizers for their 
hospitality and for the warm atmosphere provided at the workshop. This work 
has been partially supported by MIUR (Italy).


\end{document}